\documentclass[aps,prd,preprint,superscriptaddress,showpacs,nofootinbib]{revtex4}

\usepackage{graphicx}

\def\ie{{\it i.e.}}
\def\eg{{\it e.g.}}

\def\nn{\nonumber}

\def\bwt{\begin{widetext}}
\def\ewt{\end{widetext}}
\def\be{\begin{equation}}
\def\ee{\end{equation}}
\def\bea{\begin{eqnarray}}
\def\eea{\end{eqnarray}}
\def\bean{\begin{eqnarray*}}
\def\eean{\end{eqnarray*}}
\def\bary{\begin{array}}
\def\eary{\end{array}}
\def\bit{\begin{itemize}}
\def\eit{\end{itemize}}
\def\GeV{\rm GeV}

\def\su5u1{SU(5) \times U(1)}
\def\fsu5u1{SU(5) \times U(1)'}
\def\so10{SO(10)}
\def\sq20{SO(10) \times SO(10)}

\begin{document}

\title{Implications of Canonical Gauge Coupling Unification 
in High-Scale Supersymmetry Breaking}

\author{V. Barger}
\affiliation{Department of Physics, University of Wisconsin, 
Madison, WI 53706, USA}

\author{N. G. Deshpande}
\affiliation{Institute of Theoretical Science, University of Oregon, 
Eugene, OR 97403, USA}

\author{Jing Jiang}
\affiliation{Institute of Theoretical Science, University of Oregon, 
Eugene, OR 97403, USA}

\author{Paul Langacker}
\affiliation{School of Natural Sciences, Institute for Advanced Study,
  Einstein Drive, Princeton, NJ 08540, USA}

\author{Tianjun Li}

\affiliation{George P. and Cynthia W. Mitchell Institute for
Fundamental Physics, Texas A$\&$M University, College Station, TX
77843, USA }

\affiliation{Institute of Theoretical Physics, Chinese Academy of Sciences,
 Beijing 100080, P. R. China}

\date{\today}

\begin{abstract}

  We systematically construct two kinds of models with canonical gauge
  coupling unification and universal high-scale supersymmetry
  breaking. In the first we introduce standard vector-like particles
  while in the second we also include non-standard vector-like
  particles. We require that the gauge coupling unification scale is
  from $5\times 10^{15}$ GeV to the Planck scale, that the universal
  supersymmetry breaking scale is from 10 TeV to the unification
  scale, and that the masses of the vector-like particles ($M_V$) are
  universal and in the range from 200 GeV to 1 TeV.  Using two-loop
  renormalization group equation (RGE) running for the gauge couplings
  and one-loop RGE running for Yukawa couplings and the Higgs quartic
  coupling, we calculate the supersymmetry breaking scales, the gauge
  coupling unification scales, and the corresponding Higgs mass
  ranges.  When the vector-like particle masses are less than 1 TeV,
  these models can be tested at the LHC.

\end{abstract}


\preprint{MADPH-07-1477, MIFP-06-36, OITS-787, hep-ph/0701136}

\maketitle


\section{Introduction}

There is no known symmetry in effective field theory or string
theory that can constrain the cosmological constant $\Lambda_{\rm CC}$
to be zero. Why the cosmological constant is so tiny compared to
the Planck scale $M_{\rm Pl}$ or string scale $M_{\rm String}$
($\Lambda_{\rm CC} \sim 10^{-122} M_{\rm Pl}^4$) is a great mystery in
particle physics and cosmology.  In addition, because the Standard
Model (SM) Higgs boson mass is not stable against quantum
corrections, the weak scale, which is about 16 (15) order smaller than
$M_{\rm Pl}$ ($M_{\rm String}$), presents another puzzle. These are
the cosmological constant problem and gauge hierarchy problem,
respectively. Supersymmetry can solve the gauge hierarchy problem
elegantly; however, it can ameliorate but cannot solve the
cosmological constant problem.

Because there exists an enormous ``landscape'' for long-lived
metastable vacua in the Type II string theories with flux
compactifications where the moduli can be stabilized and supersymmetry
may be broken~\cite{String}, we may explain the tiny value of the
cosmological constant by the ``weak anthropic
principle''~\cite{Weinberg}, and solve the gauge hierarchy problem
simultaneously without invoking weak scale
supersymmetry~\cite{Agrawal:1998xa}.  Although the strong CP problem
is still a big challenge in the string
landscape~\cite{Donoghue:2003vs}, it can be solved by the well known
Peccei--Quinn mechanism~\cite{PQ}.  The axion solutions can be
stabilized by the gauged discrete $Z_N$ Peccei--Quinn
symmetry~\cite{Babu:2002ic,Barger:2004sf} arising from the breaking of
an anomalous $U(1)_A$ gauge symmetry in string
constructions~\cite{MGJS,Svrcek:2006yi}.  The axion can also be a cold
dark matter candidate~\cite{Barger:2004sf}.

One consequence of the string landscape is that supersymmetry can be
broken at a high scale if we have many supersymmetry breaking
parameters or many hidden sectors~\cite{HSUSY,NASD}. Because the
string landscape is mainly based on Type II orientifolds with flux
compactifications, the supersymmetry breaking soft masses and
trilinear $A$ terms are generically about $M_{\rm String}^2/M_{\rm
Pl}$, at least in the known models~\cite{Camara:2004jj,LRS}.  We shall
assume universal supersymmetry breaking in this paper.

Supposing that the cosmological constant and gauge hierarchy problems
are indeed solved in the string landscape, what would the guiding
principle for our model building and new physics search be?  In this
paper, we consider canonical gauge coupling unification as our main
guiding principle to study new physics in the extensions of the SM,
which would be expected in Grand Unified Theories (GUTs).  Achieving
the SM gauge coupling unification for high-scale supersymmetry
breaking is an interesting question. It is well known that gauge
coupling unification cannot be achieved in the SM with the canonical
normalization of the $U(1)_Y$ hypercharge interaction, \ie, the
Georgi-Glashow $SU(5)$ normalization~\cite{Langacker:1991an}, unless
we introduce additional vector-like particles at the weak
scale~\cite{Frampton:1983sh,Choudhury:2001hs,Morrissey:2003sc,Kehagias:2005vz}.
However, it can indeed be realized at about $10^{16-17}$ GeV for
non-canonical $U(1)_Y$ normalizations~\cite{Barger:2005gn}.

In this paper we systematically construct the models with canonical
gauge coupling unification and universal high-scale supersymmetry
breaking by introducing extra SM vector-like fermions at the weak
scale~\footnote{We do not consider new particles which are chiral with
respect to the SM gauge group because of the precision electroweak
constraints~\cite{Yao:2006px}.  They could, however, be chiral with
respect to additional gauge symmetries.}.  To avoid the dimension-6
proton decay problem and quantum gravity effects, we require that the
gauge coupling unification scale ($M_U$) is in the range from $5\times
10^{15}$ GeV to the Planck scale.  We also assume that the
supersymmetry breaking scale ($M_S$) can be from 10 TeV to the
unification scale.  The masses of the vector-like fermions ($M_{V}$)
could in principle be arbitrary.  However, we restrict our attention
to the case of a universal $M_V$ in the range from 200 GeV to 1 TeV.
This is motivated by simplicity and because such particles would be
observable at the LHC.  Furthermore, in some models there are
additional symmetries which require $M_V$ to be generated by the
vacuum expectation value of a Standard Model singlet field which is
tied to the electroweak scale~\cite{Cvetic:1997ky}.  To have such
gauge coupling unification, we show that the total contributions to
the one-loop beta function of $SU(2)_L$ ($\Delta b_2$) from the
vector-like fermions must be equal to those of $SU(3)_C$ ($\Delta
b_3$), \ie ~$\Delta b_2 =
\Delta b_3$, and we also obtain the constraint $2/5 \le \Delta b_2 -
\Delta b_1 \le 13/5$, where $\Delta b_1$ is the total contribution to
the one-loop beta function of $U(1)_Y$.  There are only finite
possibilities for $\Delta b_2 - \Delta b_1$ due to the quantization of
$\Delta b_i$.  To systematically study gauge coupling unification with
high-scale supersymmetry breaking, we employ the one-loop beta
function equivalent relations among the particle sets, which was
originally proposed in Ref.~\cite{LMN}. If the gauge coupling
unification can be achieved in a model with a set of vector-like
fermions which have $\Delta b_2 = \Delta b_3$ and $\Delta b_2 - \Delta
b_1=c_b$, all the models with gauge coupling unification and the
vector-like fermions which have $\Delta b_2 = \Delta b_3$ and $\Delta
b_2 - \Delta b_1=c_b$ can be constructed by adding particles such that
the one-loop beta function equivalent relations hold for the
additional particle sets.

We consider two kinds of models. For the first kind, we introduce the
standard vector-like particles whose quantum numbers are identical to
those of the SM fermions and their Hermitian conjugates, the particles
in the $SU(5)$ symmetric representation and their Hermitian
conjugates, and the $SU(5)$ adjoint particles. For the second kind, we
introduce non-standard vector-like particles which are charged under
the $SU(3)_C\times SU(2)_L$ and neutral under $U(1)_Y$. These
particles can arise from string
constructions~\cite{Dienes:1996du,Blumenhagen:2005mu}~\footnote{In
  some cases these models imply fractional electric charges, and would
  be allowed only for non-standard cosmologies.}.  After identifying
viable models, we use two-loop renormalization group equation (RGE)
running for the SM gauge couplings and one-loop RGE running for the
Yukawa couplings and the Higgs quartic coupling to calculate the
supersymmetry breaking scales, gauge coupling unification scales, and
the corresponding Higgs mass ranges for the models with simple sets of
extra vector-like fermions for $M_V=200$ GeV and 1 TeV. In the first
kind of models, $\Delta b_2 - \Delta b_1$ can only be equal to $6/5$
and $12/5$, and then the corresponding supersymmetry breaking scale
can only be around $10^{10}$ GeV and $10^{15}$ GeV, respectively.
In the second kind, $\Delta b_2 - \Delta b_1$ can be
$n/5$ with $n=2, 3, ..., 13$, and the supersymmetry breaking scale
can be from $10^{5}$ GeV to $10^{16}$ GeV if we include uncertainties
from the threshold corrections at the scales $M_V$, $M_S$ and $M_U$.
The masses of the vector-like fermions are within the reach of
the Large Hadron Collider (LHC).

We briefly discuss the phenomenological consequences
of the models, which will be presented in detail elsewhere.

This paper is organized as follows: in Section II, we present our
calculation procedure. We consider canonical gauge coupling
unification and the Higgs mass ranges in the models with standard
vector-like particles and non-standard vector-like particles in
Sections III and IV, respectively. In Section V, we comment on
phenomenological consequences.  Our discussions and conclusions are in
Section VI. The renormalization group equations are given in Appendix
A, and the two-loop beta functions for the additional vector-like
particles are given in Appendix B.

\section{Calculation Procedure}

We consider models with canonical gauge coupling unification where the
supersymmetry breaking scale is from 10 TeV to the unification scale.
In this range the constraints on the electric dipole moments (EDMs) of
the electron and neutron due to the generic CP violations in the
supersymmetry breaking soft terms can automatically be satisfied.  The
cosmological constant problem and gauge hierarchy problem are assumed
to be solved by the string landscape. We assume that the strong CP
problem can be solved by the Peccei--Quinn mechanism. The axion can be
a cold dark matter candidate.  The additional vector-like fermions
could also provide possible cold dark matter candidates.  Similar to
the new minimal SM~\cite{Davoudiasl:2004be}, the neutrino masses and
mixings can be explained by the see-saw mechanism by introducing two or
three right-handed neutrinos~\cite{Seesaw}, and the baryon asymmetry
can be generated by leptogenesis~\cite{Fukugita:1986hr} or other
mechanisms.

In supersymmetric models there generically exist one pair of Higgs
doublets $H_u$ and $H_d$.  We define the SM Higgs doublet $H$, which
is fine-tuned to have a weak-scale mass, as $H \equiv -\cos\beta i
\sigma_2 H_d^*+\sin\beta H_u$, where $\sigma_2$ is the second Pauli
matrix and $\tan\beta $ is a mixing
parameter~\cite{NASD,Barger:2004sf}.  Inspired by the supersymmetry
breaking on Type II orientifolds with flux
compactifications~\cite{Camara:2004jj,LRS}, we assume universal
supersymmetry breaking at scale $M_S$, {\it i.e.}, the gauginos,
squarks, sleptons, Higgsinos, and the other combination of the scalar
Higgs doublets ($\sin\beta i \sigma_2 H_d^*+\cos\beta H_u $) have a
universal supersymmetry breaking soft mass around $M_S$.

We require that the gauge coupling unification scale is higher than
$5\times 10^{15}$ GeV so that the dimension-6 proton decay via
exchange of the $X$ and $Y$ gauge bosons can be suppressed, and that
the scale is smaller than the Planck scale ($2.4\times 10^{18}$ GeV)
so that quantum gravity effects can be neglected~\footnote{Unification
at the string scale ($\sim 5 \times 10^{17}$ GeV) for weakly coupled
heterotic string theory \cite{Dienes:1996du} is considered in
\cite{Barger:2006fm}.}.  To achieve canonical gauge coupling
unification, we introduce vector-like fermions, and for simplicity we
assume that their masses ($M_V$) are universal and from 200 GeV to 1
TeV so that they can be observed at the LHC.  Our analysis can be
easily extended to the cases where $M_V$ takes either non-universal or
higher values.  The superpartners of these vector-like fermions
(scalar components in the supermultiplets) are assumed to have
supersymmetry breaking soft masses around $M_{S}$.  If $M_{S} \sim
M_{U}$, the canonical gauge coupling unification is realized in the SM
through the introduction of the vector-like particles.

The one-loop $\Delta b_i$ relevant from $M_{S}$ to $M_{U}$ are given
in the following Sections.  From $M_V$ to $M_{S}$, the one-loop beta
functions $\Delta b_i$ from the vector-like fermions should be 2/3 of
those for the complete supermultiplets.  The renormalization group
equations in the SM and the Minimal Supersymmetric Standard Model
(MSSM) can be found in Appendix~\ref{apdxA}.  The two-loop beta
functions from these extra vector-like fields are given in
Appendix~\ref{apdxB}.  We consider two-loop RGE running for the SM
gauge couplings and one-loop running for the Yukawa couplings and the
Higgs quartic coupling. For simplicity, we only consider the
contributions to the gauge coupling RGE running from the Yukawa
couplings of the third family of the SM fermions, \ie, the top quark,
bottom quark and $\tau$ lepton Yukawa couplings. We do not consider
the contributions to the gauge coupling RGE running from the Yukawa
couplings of the extra vector-like particles.

We denote the gauge couplings for $U(1)_Y$, $SU(2)_L$, and $SU(3)_C$
as $g_Y$, $g_2$, and $g_3$, respectively, and define $g_1\equiv {\sqrt
  {5/3}} g_Y$.  The major prediction in the models with high-scale
supersymmetry breaking is the Higgs boson
mass~\cite{Barger:2004sf,Barger:2005gn,Gogoladze:2006ps}.  We can
calculate the Higgs boson quartic coupling $\lambda$ at the
supersymmetry breaking scale $M_S$
\begin{equation}
\lambda({M_S}) = \frac{ g_2^2(M_S) + 3 g_1^2(M_S)/5}{4} \cos^2 2\beta~,
\end{equation}
and then evolve it down to the weak scale. The renormalization group
equation for the Higgs quartic coupling is also given in
Appendix~\ref{apdxA}.  Using the one-loop effective Higgs potential
with top quark radiative corrections, we calculate the Higgs boson
mass by minimizing the effective potential
\be 
V_{eff} = m_h^2 H^\dagger H - \frac{\lambda}{2!} (H^\dagger H)^2 -
\frac{3}{16\pi^2} h_t^4 (H^\dagger H)^2 \left[\log\frac{h_t^2
    (H^\dagger H)}{Q^2} - \frac{3}{2}\right]\,, 
\ee
where $m_h^2$ is the bare Higgs mass squared, $h_t$ is the top quark
Yukawa coupling, and the scale $Q$ is chosen to be at the Higgs boson
mass. We use the one-loop corrected ${\overline{MS}}$ top quark Yukawa
coupling~\cite{Arason:1991ic}, which is related to the top quark pole
mass by
\be
m_t = h_t v \left(1 + \frac{16}{3}\frac{g_3^2}{16\pi^2} -
  2\frac{h_t^2}{16\pi^2}\right)\,. 
\ee

We define $\alpha_i=g_i^2/4\pi$ and denote the $Z$ boson mass as
$M_Z$.  In the following numerical calculations, we use top quark pole
mass $m_t = 171.4 \pm 2.1~\GeV$~\cite{Brubaker:2006xn}, the strong
coupling constant $\alpha_3(M_Z) = 0.1189 \pm
0.0010$~\cite{Bethke:2006ac}.  The fine structure constant
$\alpha_{EM}$, weak mixing angle $\theta_W$ and Higgs vacuum
expectation value (VEV) $v$ at $M_Z$ are~\cite{Yao:2006px}
\bea
&&\alpha^{-1}_{EM}(M_Z) = 127.904 \pm 0.019\,, \nonumber \\
&&\sin^2\theta_W(M_Z) = 0.23122 \pm 0.00015\,, \nonumber \\
&&v = 174.10\,{\rm GeV}\,.
\eea

\section{Models with Standard Vector-like Particles}

To achieve canonical gauge coupling unification, we first introduce
the vector-like particles whose quantum numbers are the same as those
of the SM fermions and their Hermitian conjugates, particles in the
$SU(5)$ symmetric representation and their Hermitian conjugates, and
the $SU(5)$ adjoint particles. Their quantum numbers under 
$SU(3)_C \times SU(2)_L \times U(1)_Y$ and their
contributions to one-loop beta functions, $\Delta b \equiv (\Delta
b_1, \Delta b_2, \Delta b_3)$ as complete supermultiplets are
\begin{eqnarray}
&& XQ + {\overline{XQ}} = {\mathbf{(3, 2, {1\over 6}) + ({\bar 3}, 2,
-{1\over 6})}}\,, \quad \Delta b =({1\over 5}, 3, 2)\,;\\ 
&& XU + {\overline{XU}} = {\mathbf{ ({3},
1, {2\over 3}) + ({\bar 3},  1, -{2\over 3})}}\,, \quad \Delta b =
({8\over 5}, 0, 1)\,;\\ 
&& XD + {\overline{XD}} = {\mathbf{ ({3},
1, -{1\over 3}) + ({\bar 3},  1, {1\over 3})}}\,, \quad \Delta b =
({2\over 5}, 0, 1)\,;\\  
&& XL + {\overline{XL}} = {\mathbf{(1,  2, {1\over 2}) + ({1},  2,
-{1\over 2})}}\,, \quad \Delta b = ({3\over 5}, 1, 0)\,;\\ 
&& XE + {\overline{XE}} = {\mathbf{({1},  1, {1}) + ({1},  1,
-{1})}}\,, \quad \Delta b = ({6\over 5}, 0, 0)\,;\\ 
&& XG = {\mathbf{({8}, 1, 0)}}\,, \quad \Delta b = (0, 0, 3)\,;\\ 
&& XW = {\mathbf{({1}, 3, 0)}}\,, \quad \Delta b = (0, 2, 0)\,;\\
&& XT + {\overline{XT}} = {\mathbf{(1, 3, 1) + (1, 3,
-1)}}\,, \quad \Delta b =({{18}\over 5}, 4, 0)\,;\\ 
&& XS + {\overline{XS}} = {\mathbf{(6,  1, -{2\over 3}) + ({\bar 6},
1, {2\over 3})}}\,, \quad \Delta b = ({16\over 5}, 0, 5)\,;\\ 
&& XY + {\overline{XY}} = {\mathbf{(3, 2, -{5\over 6}) + ({\bar 3}, 2,
{5\over 6})}}\,, \quad \Delta b =(5, 3, 2)\,.\,
\end{eqnarray}

There are three mass scales in our models: the universal mass for the
vector-like particles $M_V$, the supersymmetry breaking scale $M_S$,
and the gauge coupling unification scale $M_U$.  The viable values of
$\Delta b$ for our choices of scales: 200 GeV $\le M_V \le$ 1 TeV, 10
TeV $\le M_S \le M_U$ and $5.0 \times 10^{15}$ GeV $< M_U <$ $2.4
\times 10^{18}$ GeV, are limited.  At one-loop level, only the
relative differences between the beta functions are relevant so that
$(\Delta b_1, \Delta b_2, \Delta b_3)$ is essentially equivalent to
$(0, \Delta b_2 - \Delta b_1, \Delta b_3 - \Delta b_1)$, \ie, increasing
or decreasing $\Delta b_1$, $\Delta b_2$, and $\Delta b_3$ by the same
amount does not affect these mass scales, but they do increase or
decrease the strength of the unified gauge couplings, respectively.
As long as we keep $\Delta b_1$ less than around $10$, the gauge
couplings at the unification scale will remain perturbative.

\begin{figure}[t]
\centering
\includegraphics[width=8cm]{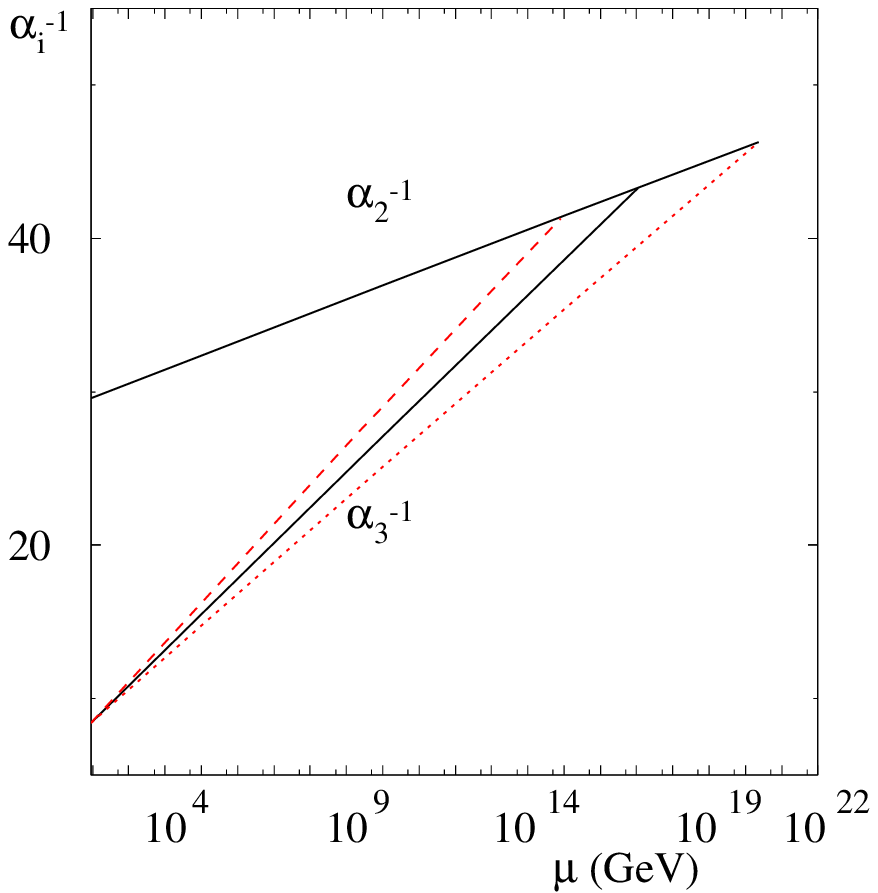}
\includegraphics[width=8cm]{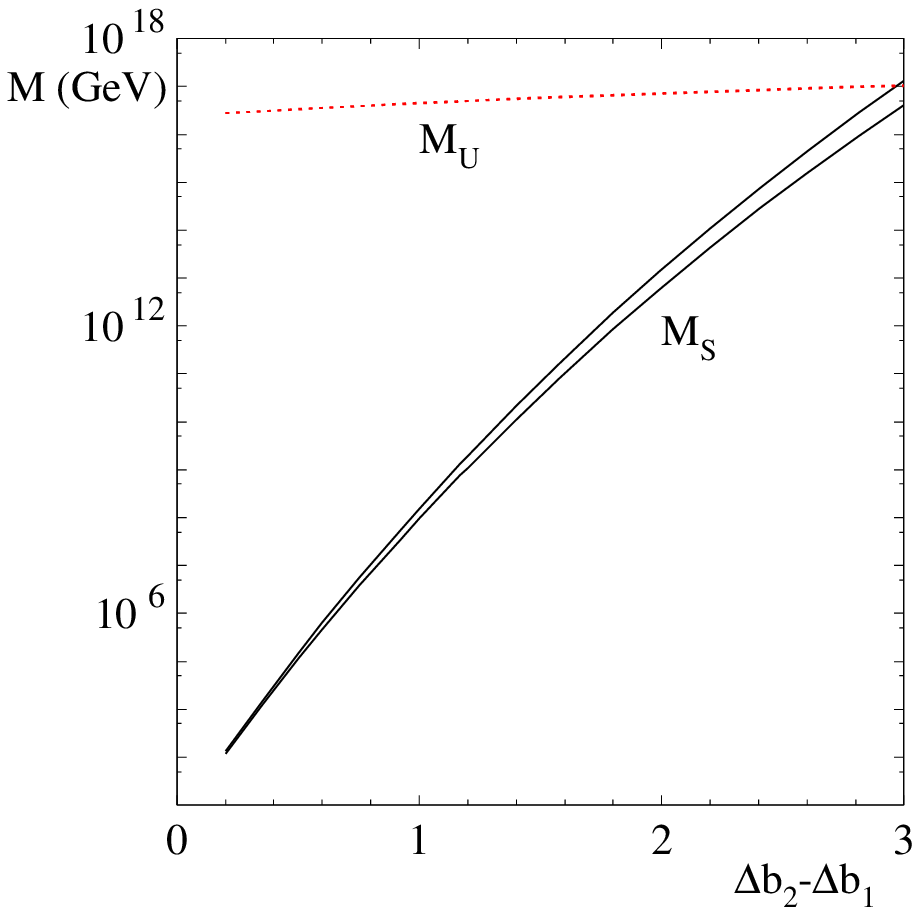}
\caption{Left: The intersections of the upper solid line with the
  dashed line, solid line and the dotted line show the one-loop
  $SU(3)_C \times SU(2)_L$ gauge coupling unification scale for
  $(\Delta b_2 = 1, \Delta b_3 = 0)$, $(\Delta b_2 = 1, \Delta b_3 =
  1)$, and $(\Delta b_2 = 1, \Delta b_3 = 2)$, respectively.  Right:
  Mass scales $M_S$ (solid) and $M_U$ (dotted) as functions of $\Delta
  b_2 - \Delta b_1$ from one-loop RGE running.  The upper curves
  correspond to $M_V = 200~\GeV$ and the lower curves $M_V = 1~{\rm
    TeV}$.}
\label{fig:oneloop01}
\end{figure}

Let us first study the possible values for $\Delta b_3 - \Delta b_2$.
The choices of $\Delta b_3 - \Delta b_2 \le -1$ and $\Delta b_3 -
\Delta b_2 \ge 1$ respectively produce too small and too large values
for the $SU(2)_L \times SU(3)_C$ gauge coupling unification scale
$M_U$.  Assuming $M_S=M_U$ and the SM gauge couplings at the weak
scale, we show the one-loop $SU(2)_L\times SU(3)_C$ unification scale
$M_U$ for the cases $(\Delta b_2 = 1, \Delta b_3 = 0)$ and $(\Delta
b_2 = 1, \Delta b_3 = 2)$ in the left plot of
Fig.~\ref{fig:oneloop01}.  For $(\Delta b_2 = 1, \Delta b_3 = 0)$,
$M_U$ is smaller than $8 \times 10^{14}~\GeV$, which is the maximal
unification scale for the case $\Delta b_3 - \Delta b_2 \le -1$.  For
$(\Delta b_2 = 1, \Delta b_3 = 2)$, $M_U$ is larger than
$10^{20}~\GeV$, which is the minimal unification scale for $\Delta b_3
- \Delta b_2 \ge 1$.  Because the two-loop RGE running can only change
the unification scale by a factor less than 5 for the models we have
studied in this paper and $\Delta b_2$ and $\Delta b_3$ only take
integer values, we obtain~\footnote{The argument becomes even stronger
for $M_S < M_U$, with $M_U$ becoming even smaller or larger for
$\Delta b_3 - \Delta b_2 = \mp 1$, respectively.  On the other hand
the argument would be weakened if we allowed $M_V$ much larger than 1
TeV, \ie, in that case $\Delta b_3 \ne \Delta b_2$ would be allowed.} 
that $\Delta b_3 = \Delta b_2 $.  We also observe that gauge coupling
unification including a canonically normalized $U(1)_Y$ requires $2/5
\le \Delta b_2 - \Delta b_1 \le 13/5$ in the models with high-scale
supersymmetry breaking.  For $\Delta b_2 - \Delta b_1 > 3$, $M_S$ is
larger than $M_U$, and $1/5$, $14/5$ or $3$ cannot be generated from
the given particle sets.

In the right plot of Fig.~\ref{fig:oneloop01} we show the dependence
of $M_S$ and $M_U$ on $\Delta b_2 -\Delta b_1$, based on one-loop RGE
running for the SM gauge couplings.  In two-loop RGE running, the
actual values of $\Delta b$'s will shift $M_S$ and $M_U$ away from
those indicated by the curves.  Curves for both $M_S$ and $M_U$ are
plotted for $M_V = 200~\GeV$ and $M_V = 1~{\rm TeV}$.  However,
for $M_U$ the two dotted curves are too close to each other to be
discerned.  The solid curves are for $M_S$, with the upper one for $M_V =
200~\GeV$ and the lower for $M_V = 1~{\rm TeV}$.  As we increase
$\Delta b_2 -\Delta b_1$, the increase in $M_U$ is gradual, but the
increase in $M_S$ is very rapid.

Using the constraints on $\Delta b_1$, $\Delta b_2$, and $\Delta b_3$,
we are ready to generate the complete sets of vector-like particles that
will ensure canonical gauge coupling unification. Because $\Delta b_3
= \Delta b_2 $ and $2/5 \le \Delta b_2 - \Delta b_1 \le 13/5$, there
are only finite possibilities for $\Delta b_2 - \Delta b_1$ due to the
quantization of $\Delta b_i$.  We employ the equivalent relations of
the one-loop beta function for the particle sets~\cite{LMN}. If we can
achieve canonical gauge coupling unification by introducing one set of
the vector-like fermions with $\Delta b_3 = \Delta b_2 $ and $\Delta
b_2 - \Delta b_1=c_b$, it also holds for one-loop equivalent sets,
defined as those with the same $\Delta b_3 - \Delta b_2 $ and $\Delta
b_2 - \Delta b_1$ at one loop, because the two-loop effects give only
small corrections.  The complete independent one-loop beta function
equivalent relations for the particle sets are~\cite{LMN}
\begin{eqnarray}
\label{eq:eqv1}
&& EQV1: XQ +  XU  + XE  \sim 0 ~,~~{\rm or}~~
{\overline{XQ}} + {\overline{XU}} +  {\overline{XE}}  \sim 0 \,; \\
%
%
&& EQV2: XD + XL  \sim 0 ~,~~{\rm or}~~{\overline{XD}}+
{\overline{XL}} \sim 0 \,; \\ 
%
&& EQV3: XW +XG + XY + {\overline{XY}} \sim 0 \,; \\
\label{eq:eqv4}
&& EQV4: XQ+XT+XS\sim 0 ~,~~{\rm or}~~
{\overline{XQ}} + {\overline{XT}} +  {\overline{XS}}  \sim 0 \,; \\
&& EQV5: 2 (XD + {\overline{XD}}) + XE + {\overline{XE}} +  XW \sim 0 \,; \\
&& EQV6: XL + {\overline{XL}}+ 2(XE + {\overline{XE}}) + XW +XG \sim 0 \,; \\
&& EQV7: XD +  XE \sim XU   ~,~~{\rm or}~~
{\overline{XD}} + {\overline{XE}} \sim {\overline{XU}} \,; \\
%
&& EQV8: 3(XE + {\overline{XE}}) + 2 XW \sim XT + {\overline{XT}} \,; \\
&& EQV9: XU  + XE  \sim 2 XD  + XY ~,~~{\rm or}~~{\overline{XU}} +
{\overline{XE}} 
\sim 2  {\overline{XD}}+  {\overline{XY}} \,.~\,
\end{eqnarray}
where $0$ means the zero particle set.  Equivalent relations
(\ref{eq:eqv1}) -- (\ref{eq:eqv4}) correspond respectively to 10, 5, 24,
and 15-plets of $SU(5)$.

The conditions $\Delta b_2 = \Delta b_3$ and $2/5 \le \Delta b_2 -
\Delta b_3 \le 13/5$ for canonical gauge coupling unification are satisfied by
the simple sets
\begin{eqnarray}
&&Z1: XW+2(XD+{\overline{XD}})\,, ~~\Delta b = ({4\over5}, 2, 2) \sim (0,
{6\over5}, {6\over5})\,; \\
&&Z2: XW+3(XD+{\overline{XD}})+(XL+{\overline{XL}})
\,, ~~\Delta b = ({9\over5}, 3, 3) \sim (0,
{6\over5}, {6\over5})\,; \\
&&Z3: XQ+ {\overline{XQ}}+XU + {\overline{XU}}\,, ~~\Delta
b =({9\over5}, 3, 3) \sim (0, {6\over5}, {6\over5})\,; \\
&&Z4: XQ+ {\overline{XQ}}+XD + {\overline{XD}}
+ XE + {\overline{XE}} \,, ~~\Delta
b =({9\over5}, 3, 3) \sim (0, {6\over5}, {6\over5})\,; \\
&&Z5: XG+3(XL+{\overline{XL}})\,, ~~\Delta b =({9\over5}, 3, 3) \sim (0,
{6\over5}, {6\over5})\,; \\
&&Z6: XG+XW+XL+ {\overline{XL}} + XE+{\overline{XE}} \,, ~~\Delta b
=({9\over 5}, 
3, 3) \sim (0, {6\over5}, {6\over5})\,; \\
&&Z7: XG+XW+ XD+{\overline{XD}} +
2 (XL+ {\overline{XL}}) + XE+{\overline{XE}}\,,  \nonumber \\
&&~~~~~~~~~~~~\Delta b =({14\over 5},
4, 4) \sim (0, {6\over5}, {6\over5})\,; \\
&&Z8: XQ+ {\overline{XQ}}+XD + {\overline{XD}}\,, ~~\Delta
b =({3\over 5}, 3, 3) \sim (0,{12\over5}, {12\over5})\,; \\
&&Z9: XG+XW+XL+ {\overline{XL}}\,, ~~\Delta b =({3\over 5},
3, 3) \sim (0, {12\over5}, {12\over5})\,; \\
&&Z10: XT + {\overline{XT}} + XW + 2 XG\,, ~~\Delta
b =({18\over 5}, 6, 6) \sim (0,{12\over5}, {12\over5})\,,
\end{eqnarray}
all of which either have $\Delta b_2 - \Delta b_1=6/5$ (sets $Zi$, $i
= 1, \dots, 9$) or $\Delta b_2 - \Delta b_1=12/5$ (sets $Z8$, $Z9$,
$Z10$). The sets $Zi$ ($i=2, ..., 7$) may be generated from $Z_1$, and
$Z9$ and $Z10$ from $Z8$ by using one-loop beta function equivalent
relations:
\begin{table}[htb]
\begin{center}
\begin{tabular}{|c|c|c|c|}
\hline
Equiv. sets & Equiv. relations & Equiv. sets & Equiv. relations \\
\hline
$Z2 \sim Z1$ & 2 & $Z3 \sim Z1$ & 1, 5 \\
$Z4 \sim Z3$ & 7 & $Z5 \sim Z1$ & 2, 5, 6 \\
$Z6 \sim Z1$ & 5, 6 & $Z7 \sim Z6$ & 2 \\
$Z9 \sim Z8$ & 1, 6, 7 & $Z10 \sim Z8$ & 5, 6, 8 \\
\hline
\end{tabular}
\end{center}
\caption{Equivalent sets and the equivalent relations involved.
  }
\label{tbl:eqicon}
\end{table}

With two-loop RGE running for the gauge couplings and one-loop running
for the Yukawa and Higgs quartic coupling, we show the supersymmetry
breaking scales, the gauge coupling unification scales and the
corresponding Higgs mass ranges for $M_V = 200~\GeV$ and 1 TeV in
Table~\ref{tbl:twoloop1}.  The Higgs boson mass ranges correspond to
the variation of $\tan\beta$ between 1.5 and 50, with smaller
$\tan\beta$ giving a smaller Higgs boson mass, and $\alpha_s$ and
$m_t$ with their $1\sigma$ ranges.  For the same $\Delta b_2 - \Delta
b_1$, the actual values of $\Delta b_1$ and $\Delta b_2$, as well as
the different two-loop beta functions due to the different additional
particle contents can affect RGE running, and hence these mass scales
and Higgs boson masses.  This is evident in comparing the $Z1$, $Z3$
and $Z5$ sets.  The $Z1$ set has different $\Delta b_1$ from $Z3$ and
$Z5$, while $Z3$ and $Z5$ differ in the two-loop beta functions due to
the different extra particles involved.  For $Z1$ through $Z9$ the
Higgs mass ranges are from about 119~GeV to 143 GeV for $M_V=200~\GeV$
and from about 122~GeV to 145~GeV for $M_V=1~{\rm TeV}$. The Higgs
mass ranges are from 103~GeV to 143~GeV and from 113~GeV to 145~GeV in
the model with the $Z10$ set for $M_V=200~\GeV$ and $M_V=1~{\rm TeV}$,
respectively.  The Higgs mass ranges are larger (\ie, a lighter Higgs
is allowed) in the model with the $Z10$ set than the other models
because the values of $\Delta b_1$ and $\Delta b_2$ are larger.  In
general, for the models with $\Delta b_2 - \Delta b_1 = 6/5$, the
supersymmetry breaking scale is around $10^{10}~\GeV$.  For those with
$\Delta b_2 - \Delta b_1 = 12/5$, the supersymmetry breaking scale is
about $10^{15}~\GeV$, which can be considered as the GUT scale up to
uncertainties from the threshold corrections at the scales $M_V$,
$M_S$, and $M_U$.  For a particular model with the $Zi$ set, the SM
gauge couplings, the Higgs quartic coupling at the supersymmetry
breaking scale, as well as the physical Higgs mass will decrease if we
increase $M_V$.

\begin{table}[htb]
\begin{center}
\begin{tabular}{|c|c|c|c|c|c|c|c|}
\hline
\multicolumn{2}{|c|}{} &\multicolumn{3}{c|}{$M_V =
200~\GeV$}&\multicolumn{3}{c|}{$M_V = 1000~\GeV$} \\
\cline{1-8}
$Z$  & $\Delta b_2 - \Delta b_1$ & $M_S$  & $M_U$ & $m_h$
& $M_S$ & $M_U$ & $m_h$ \\
\hline
$Z1$ & 6/5 & $2.9 \times 10^{10}$ & $3.6 \times 10^{16}$ & 123 - 144&
$1.8 \times 10^{10}$ & $3.3 \times 10^{16}$ & 125 - 145 \\
$Z2$ & 6/5 & $2.5 \times 10^{10}$ & $4.4 \times 10^{16}$ & 121 - 143 &
$1.5 \times 10^{10}$ & $4.0 \times 10^{16}$ & 124 - 145 \\
$Z3$ & 6/5 & $4.0 \times 10^{10}$ & $4.0 \times 10^{16}$ & 121 - 143 &
$2.3 \times 10^{10}$ & $3.7 \times 10^{16}$ & 124 - 145 \\
$Z4$ & 6/5 & $5.1 \times 10^{10}$ & $4.0 \times 10^{17}$ & 121 - 143 &
$2.9 \times 10^{10}$ & $3.7 \times 10^{16}$ & 124 - 145 \\
$Z5$ & 6/5 & $6.6 \times 10^{10}$ & $8.1 \times 10^{16}$ & 121 - 143 &
$4.5 \times 10^{10}$ & $7.1 \times 10^{16}$ & 123 - 145 \\
$Z6$ & 6/5 & $1.3 \times 10^{10}$ & $7.0 \times 10^{16}$ & 121 - 143 &
$8.2 \times 10^{10}$ & $6.1 \times 10^{16}$ & 123 - 145 \\
$Z7$ & 6/5 & $9.6 \times 10^{10}$ & $9.8 \times 10^{16}$ & 119 - 143 &
$6.7 \times 10^{10}$ & $8.1 \times 10^{16}$ & 122 - 145 \\
\hline
$Z8$ & 12/5 & $8.2 \times 10^{15}$ & $6.2 \times 10^{16}$ & 119 - 143 &
$3.0 \times 10^{15}$ & $5.6 \times 10^{16}$ & 123 - 145 \\
$Z9$ & 12/5 & $3.9 \times 10^{15}$ & $9.6 \times 10^{16}$ & 119 - 143 &
$1.6 \times 10^{15}$ & $8.3 \times 10^{16}$ & 123 - 145 \\
$Z10$ & 12/5 & $3.0 \times 10^{15}$ & $3.1 \times 10^{17}$ & 103 - 143 &
$1.4 \times 10^{15}$ & $2.1 \times 10^{17}$ & 113 - 145 \\
\hline
\end{tabular}
\end{center}
\caption{
  The supersymmetry breaking scales, the gauge coupling unification scales,
  and the corresponding Higgs mass ranges for $M_V = 200~\GeV$
  and $1000~\GeV$ in the models with $Zi$ sets of vector-like particles.
  All masses are in GeV.}
\label{tbl:twoloop1}
\end{table}

\begin{figure}[htb]
\centering
\includegraphics[width=8cm]{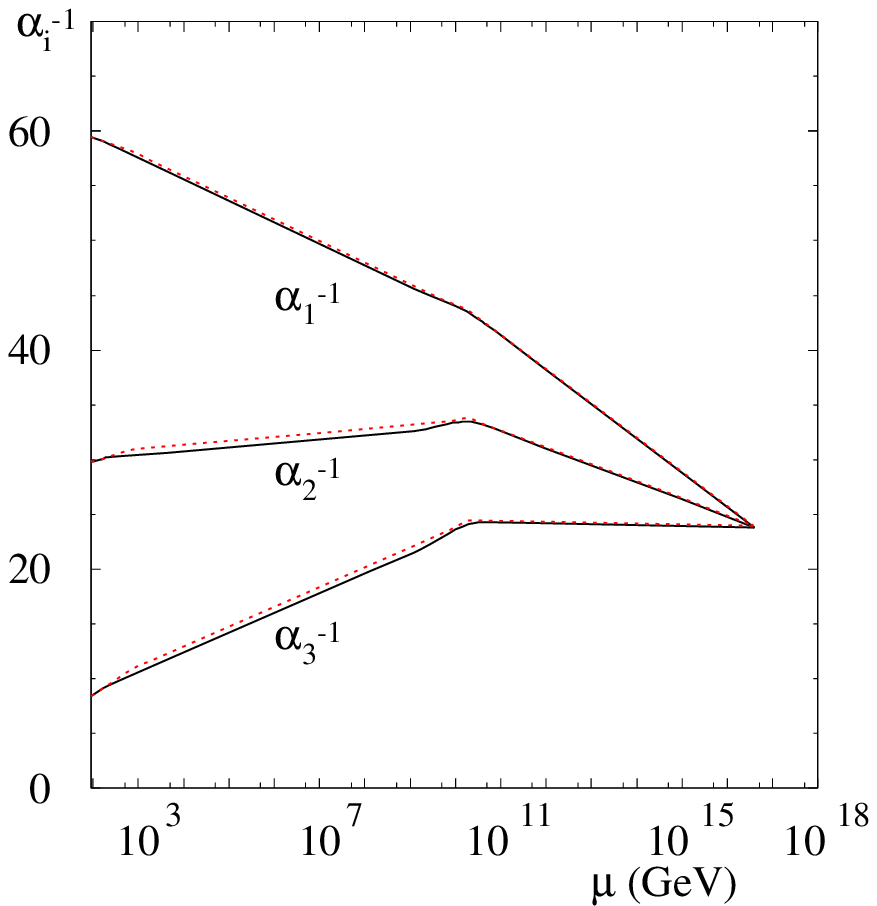}
\caption{Two-loop gauge coupling unification in the model with the
  $Z3$ set of vector-like particles.  The solid curves are for
  $M_V=200~\GeV$, while the nearly overlapping dotted curves are for
  $M_V=1~{\rm TeV}$.}
\label{fig:twoloopZ2}
\end{figure}

As an example, we show the two-loop RGE running for the SM gauge
couplings in the model with the $Z3$ set in Fig.~\ref{fig:twoloopZ2}.

\section{Models with Non-Standard Vector-like Particles}

In string model building, we may have vector-like particles which are
charged under $SU(3)_C\times SU(2)_L$ and neutral under
$U(1)_Y$~\cite{Dienes:1996du,Blumenhagen:2005mu}.  Often, such
particles also carry hidden sector charges.  Thus, we introduce such
non-standard vector-like particles in this Section. Their quantum
numbers under $SU(3)_C \times SU(2)_L \times U(1)_Y$ and their
contributions to one-loop beta functions as complete supermultiplets
are
\begin{eqnarray}
\label{eq:q0}
&& XQ0 + {\overline{XQ0}} = {\mathbf{(3, 2, 0) + ({\bar 3}, 2,0)}}\,, 
\quad \Delta b =(0, 3, 2)\,;\\   
&& XD0 + {\overline{XD0}} = {\mathbf{({\bar 3},  1, 0) + ({3},
1, 0)}}\,, \quad \Delta b = (0, 0, 1)\,;\\ 
\label{eq:l0}
&& XL0 + {\overline{XL0}} = {\mathbf{(1,  2, 0) + ({1},  2,0)}}\,, 
\quad \Delta b = (0, 1, 0) \,;\\
&& XS0 + {\overline{XS0}} = {\mathbf{(6,  1, 0) + ({\bar 6},
1, 0)}}\,, \quad \Delta b = (0, 0, 5) ~.~\,
\end{eqnarray}
We do not consider $XU0 + \overline{XU0}$, $XT0$ or $XY0 +
\overline{XY0}$ because they are equivalent to $XD0 + \overline{XD0}$,
$XW$, and $XQ0 + \overline{XQ0}$, respectively.

Note that the states in (\ref{eq:q0}) and (\ref{eq:l0}) have
half-integer electric charge, and that the lightest such particles
would be stable. Due to the stringent experimental limits on the
natural abundances of such particles and their bound
states~\cite{Lee:2002sa}, they would have to be much more massive than
the reheating temperature after a period of
inflation~\cite{Kudo:2001ie}. Thus, for them to exist at the TeV scale
the reheating temperature would have to be extremely
low~\cite{Gogoladze:2006si}.

The additional independent one-loop beta function
equivalent relations for the standard and non-standard particle sets are
\begin{eqnarray}
&& NEQV1:  2 XD0 + 3 XL0 \sim XQ0 ~,~~{\rm or}~~
2 {\overline{XD0}} + 3 {\overline{XL0}} \sim {\overline{XQ0}} \,; \\
&& NEQV2: 5 XD0 \sim XS0 ~,~~{\rm or}~~
5 {\overline{XD0}} \sim {\overline{XS0}} \,; \\
&& NEQV3:  2 XD + XE + 2 XL0  \sim 0 ~,~~{\rm or}~~
 2 {\overline{XD}} + {\overline{XE}} + 2 {\overline{XL0}} \sim 0 \,; \\
&& NEQV4: 5 XE + 6 XD0+ 6 XL0 \sim 0 ~,~~{\rm or}~~
5 {\overline{XE}} + 6 {\overline{XD0}} +
6 {\overline{XL0}}  \sim 0 \,.~\,
\end{eqnarray}

We consider the following simple sets
of standard and non-standard vector-like particles
\begin{eqnarray}
&&NZ1: XD0+{\overline{XD0}} + XL+{\overline{XL}} \,, ~~\Delta b =
({3\over5}, 1, 1) \sim (0, {2\over5}, {2\over5})\,; \\ 
&&NZ2: 4(XD0+{\overline{XD0}}) + XT + {\overline{XT}}\,, ~~\Delta b =
({18\over5}, 4, 4) \sim (0, {2\over5}, {2\over5})\,; \\ 
&&NZ3: XL0+{\overline{XL0}} +  XD+{\overline{XD}}\,, ~~\Delta b =
({2\over5}, 1, 1) \sim (0, {3\over5}, {3\over5})\,; \\ 
&&NZ4: 2(XD0+{\overline{XD0}}) +  XW + XE+{\overline{XE}}\,, ~~\Delta b =
({6\over5}, 2, 2) \sim (0, {4\over5}, {4\over5})\,; \\ 
&&NZ5: XD0+ {\overline{XD0}}+ XL0 + {\overline{XL0}}\,,
~~~\Delta b =(0, 1, 1)\,; \\ 
&&NZ6: XQ0 + {\overline{XQ0}}+XU+ {\overline{XU}}\,, ~~\Delta b =({8\over 5},
3, 3) \sim (0, {7\over5}, {7\over5})\,; \\
&&NZ7: XQ0 + {\overline{XQ0}}+XD+ {\overline{XD}}
+ XE+ {\overline{XE}}
\,, ~~\Delta b =({8\over 5},
3, 3) \sim (0, {7\over5}, {7\over5})\,; \\
&&NZ8: XD0+{\overline{XD0}} + XD+{\overline{XD}} + XW\,, ~~\Delta b =
({2\over5}, 2, 2) \sim (0, {8\over5}, {8\over5})\,; \\ 
&&NZ9: XL0+{\overline{XL0}} + XG +2(XL+{\overline{XL}})\,, ~~\Delta b =
({6\over5}, 3, 3) \sim (0, {9\over5}, {9\over5})\,; \\ 
&&NZ10: 2(XD0+{\overline{XD0}}) + XW\,, ~~\Delta b =
(0, 2, 2) \,; \\ 
&&NZ11: XS0 + 5(XL+{\overline{XL}})\,, ~~\Delta b =
(3, 5, 5) \sim (0, 2, 2)\,; \\
&&NZ12: XD0+{\overline{XD0}} + XL0+{\overline{XL0}}+ XQ +
{\overline{XQ}}+ XU+{\overline{XU}}\,, \nn \\ 
&& \quad \quad \quad~~\Delta b =   
({9\over5}, 4, 4) \sim (0, {11\over5}, {11\over5})\,; \\ 
&&NZ13: 2(XL0+{\overline{XL0}}) + XG + XL+{\overline{XL}}\,, ~~\Delta b =
({3\over5}, 3, 3) \sim (0, {12\over5}, {12\over5})\,; \\ 
&&NZ14: 2(XL0+{\overline{XL0}}) + XT + {\overline{XT}} + 2 XG\,, ~~\Delta b =
({18\over5}, 6, 6) \sim (0, {12\over5}, {12\over5})\,; \\
&&NZ15: XQ0 + {\overline{XQ0}} + XD + {\overline{XD}} \,, ~~\Delta b
=({2\over5}, 3, 3) \sim (0, {13\over5}, {13\over5})\,.
\end{eqnarray}
%

\begin{table}[htb]
\begin{center}
\begin{tabular}{|c|c|c|c|c|c|c|c|}
\hline
\multicolumn{2}{|c|}{} &\multicolumn{3}{c|}{$M_V =
200~\GeV$}&\multicolumn{3}{c|}{$M_V = 1000~\GeV$} \\
\cline{1-8}
$NZ$  & $\Delta b_2 - \Delta b_1$ & $M_S$  & $M_U$ & $m_h$
& $M_S$ & $M_U$ & $m_h$ \\
\hline
$NZ1$ & 2/5 & $5.7 \times 10^{5}$ & $2.4 \times 10^{16}$ & 114 - 139 &
$4.8 \times 10^{5}$ & $2.3 \times 10^{16}$ & 114 - 139 \\
$NZ2$ & 2/5 & $1.8 \times 10^{6}$ & $2.9 \times 10^{16}$ & 114 - 140 &
$1.6 \times 10^{6}$ & $6.4 \times 10^{16}$ & 107 - 139 \\
\hline
$NZ3$ & 3/5 & $9.0 \times 10^{6}$ & $2.6 \times 10^{16}$ & 119 - 142 &
$1.4 \times 10^{6}$ & $3.6 \times 10^{16}$ & 115 - 144 \\
\hline
$NZ4$ & 4/5 & $2.6 \times 10^{8}$ & $3.0 \times 10^{16}$ & 121 - 143 &
$1.8 \times 10^{8}$ & $2.8 \times 10^{16}$ & 122 - 144 \\
\hline
$NZ5$ & 1 &$1.9 \times 10^{9}$ & $3.1 \times 10^{16}$ & 124 - 144 & $1.2
\times 10^{9}$ & $3.0 \times 10^{16}$ & 125 - 145 \\
\hline
$NZ6$ & 7/5 & $3.9 \times 10^{11}$ & $4.3 \times 10^{16}$ & 121 - 144 &
$2.1 \times 10^{11}$ & $4.0 \times 10^{16}$ & 124 - 145 \\
$NZ7$ & 7/5 & $4.9 \times 10^{11}$ & $4.3 \times 10^{16}$ & 121 - 144 &
$2.5 \times 10^{11}$ & $4.0 \times 10^{16}$ & 124 - 145 \\
\hline
$NZ8$ & 8/5 & $2.7 \times 10^{12}$ & $4.2 \times 10^{16}$ & 124 - 144 &
$1.3 \times 10^{11}$ & $3.9 \times 10^{16}$ & 126 - 146 \\
\hline
$NZ9$ & 9/5 & $7.3 \times 10^{12}$ & $9.1 \times 10^{16}$ & 121 - 143 &
$3.8 \times 10^{12}$ & $7.9 \times 10^{16}$ & 124 - 145 \\
\hline
$NZ10$ & 2 & $1.5 \times 10^{14}$ & $4.8 \times 10^{16}$ & 124 - 144 &
$6.5 \times 10^{13}$ & $4.5 \times 10^{16}$ & 126 - 146 \\
$NZ11$ & 2 & $2.2 \times 10^{13}$ & $3.1 \times 10^{17}$ & 113 - 142 &
$1.2 \times 10^{13}$ & $2.3 \times 10^{17}$ & 119 - 145 \\
\hline
$NZ12$ & 11/5 & $1.2 \times 10^{15}$ & $7.4 \times 10^{16}$ & 116 - 143 &
$4.7 \times 10^{14}$ & $6.5 \times 10^{16}$ & 121 - 145 \\
\hline
$NZ13$ & 12/5 & $2.9 \times 10^{15}$ & $1.1 \times 10^{17}$ & 120 - 143 &
$1.2 \times 10^{15}$ & $9.1 \times 10^{16}$ & 123 - 145 \\
$NZ14$ & 12/5 & $2.0 \times 10^{15}$ & $3.6 \times 10^{17}$ & 104 - 143
& $9.8 \times 10^{14}$ & $2.4 \times 10^{17}$ & 113 - 145 \\
\hline
$NZ15$ & 13/5 & $4.7 \times 10^{16}$ & $6.6 \times 10^{16}$ & 118 - 143
& $1.6 \times 10^{16}$ & $6.0 \times 10^{16}$ & 123 - 145 \\
\hline
\end{tabular}
\end{center}
\caption{
 Same as Table~\ref{tbl:twoloop1}, only for the $NZi$ sets.}
\label{tbl:twoloop2}
\end{table}

With the non-standard vector-like particles, $\Delta b_2 - \Delta b_1$
can be $n/5$, where $n= 2, 3, ...,13$.  The sets with $\Delta b_3 =
\Delta b_2$ and $\Delta b_2 - \Delta b_1=6/5$ are given in the
previous Section.  Simple estimates of the supersymmetry breaking
scales and the unification scales from one-loop RGE running are
already presented in the right plot of Fig.~\ref{fig:oneloop01}.  With
two-loop RGE running for the SM gauge couplings and one-loop running
for the Yukawa couplings and Higgs quartic coupling included, we
present the more reliable supersymmetry breaking scales, gauge
coupling unification scales and the corresponding Higgs boson mass
ranges in Table~\ref{tbl:twoloop2}. The supersymmetry
breaking scales can be from $10^5~\GeV$ to the GUT scale if we include
the uncertainties from threshold corrections at $M_V$,
$M_S$ and $M_U$. In general, the supersymmetry breaking scale will be
higher for the models with larger $\Delta b_2 - \Delta b_1$.

\begin{figure}[b]
\centering
\includegraphics[width=8cm]{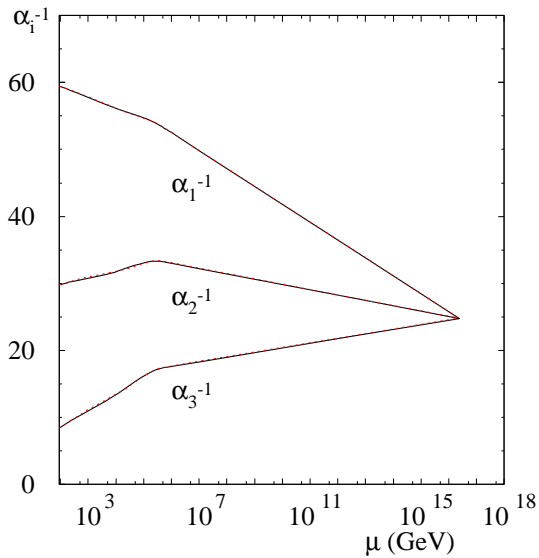}
\caption{Same as Fig.~\ref{fig:twoloopZ2}, only for the $NZ1$ set.}
\label{fig:twoloopZ7}
\end{figure}

We show the two-loop RGE running of the SM gauge couplings in the
model with the $NZ1$ set in Fig.~\ref{fig:twoloopZ7}.  Because of the
smaller $\Delta b_2 - \Delta b_1$ value, the supersymmetry breaking
scale is lower compared to Fig.~\ref{fig:twoloopZ2}.

\section{Comments on the Phenomenological Consequences}

We now address the problem of how vector-like particles have masses at
the electroweak scale. Since mass terms of vector-like particles are
invariant under the Standard Model gauge group, we are allowed to
write terms like $\overline{XQ}\,XQ$ in the Lagrangian, and the
natural scale of this mass might be the unification scale.  This would
then lead to a new fine tuning problem.  A natural mechanism to forbid
such mass terms is to embed the particles in a larger symmetry group
such that the only mass terms allowed are through Yukawa couplings
with a singlet field $S$, for example $S\, \overline{XQ}\,XQ$, with
$S$ having a VEV at the electroweak scale.  This is the mechanism of
mass generation of vector-like down-type quarks based on the group
$E_6$ which could arise from heterotic string compactification.  The
question of why vector-like particles do not occur in complete GUT
multiplets can be understood by breaking the GUT symmetry via Wilson
lines~\cite{Braun:2005ux} or orbifold projections~\cite{Orbifold}.
Another consequence of the singlet field $S$ is that we can obtain a
strong first order electroweak phase transition with the presence of
the trilinear interaction $SH^\dagger H$ in the Higgs potential,
similar to the next to the Minimal Supersymmetric Standard
Model~\cite{Pietroni,DFMHS} and the supersymmetric $U(1)'$
model~\cite{Kang:2004pp}.

The vector-like fermions can yield rich low energy phenomenology.
Models with $XD$ and $\overline{XD}$ have received a lot of attention
because they naturally occur in heterotic string inspired $E_6$
models~\cite{Barger:1985nq}.  It is interesting to note that
transformation under the Standard Model gauge group does not uniquely
specify all the properties of such vector-like particles.  For
example, the superpotential of $XD$ and $\overline{XD}$ has to be
defined before a complete description can be given.  Three
possibilities depending on lepton and baryon number assignments are
(a) down type quark, (b) leptoquark, and (c) diquark.  For a review of
the production and decays of these particles see
Ref.~\cite{Hewett:1988xc, KLN}.  They may also be quasi-stable
decaying only by higher-dimensional operators~\cite{Fairbairn:2006gg}
with cosmological and collider implications~\cite{ KLN,
Fairbairn:2006gg}.  For the models with $XU$ and ${\overline{XU}}$,
there are new effects in top and charm quark (\eg, $D$ meson) physics,
while for the models with $XD$ and ${\overline{XD}}$, we have new
effects in $B$
physics~\cite{Morrissey:2003sc,Barger:1995dd,Deshpande:2003nx}.  Also,
models with $XQ$, ${\overline{XQ}}$, $XD$ and ${\overline{XD}}$ can
explain the bottom quark forward-backward asymmetry
($A_{FB}^b$)~\cite{Choudhury:2001hs}.  Neutrino masses and mixings can
be generated if there exist $XW$ and $XL$/$\overline{XL}$, or two
$XW$, or two $XL$/$\overline{XL}$. The neutral component of $XW$ or
$XL/{\overline{XL}}$ can be a cold dark matter candidate if there
exists a discrete symmetry and their masses are around the TeV
scale~\cite{REST}. The models with $XW$, $XL$, and ${\overline{XL}}$,
may not only explain the dark matter but also generate the baryon
asymmetry via electroweak baryogenesis~\cite{Carena:2004ha}.  Similar
to split supersymmetry, the supersymmetry breaking scale may not be
higher than $10^{12}~\GeV$ in the models with $XG$ and the standard
vector-like quarks because $XG$ cannot decay fast enough via Yukawa
couplings in the superpotential to satisfy cosmological
constraints~\cite{Arvanitaki:2005fa}.  For models with $XG$ but no
other standard vector-like quarks, the cosmological constraint on $XG$
and the phenomenological consequences deserve detailed study because
$XG$ can be stable at least in some orbifold models.  Similarly,
whether the non-standard vector-like particles can decay, and the
cosmological constraints on the non-standard vector-like particles and
their phenomenological consequences deserve further detailed study.

Let us focus on the experimentally viable models which have standard
vector-like particles.  Suppose that the axion is the cold dark matter
candidate, and we introduce two or three right-handed neutrinos to
explain the neutrino masses and mixings and the baryon asymmetry. The
simple models with $10^{10}$ GeV-scale supersymmetry breaking are
those with $Z1$ and $Z3$ sets, and the simplest with $10^{15}$
GeV-scale supersymmetry breaking is the one with the $Z8$ set.  If
that axion does not contribute to the dominant cold dark matter
density, and the neutrino masses and mixings are generated due to the
$R-$parity violating terms~\cite{Grossman:2003gq}, the model with the
$Z2$ set is the simplest which has a dark matter candidate and can
explain the baryon asymmetry.

The phenomenological consequences of our models, for example, new
effects in the meson physics, CP violation, and the collider
signatures at the LHC will be presented in detail elsewhere.

\section{Discussions and Conclusions}

We studied the canonical gauge coupling unification in the extensions
of the SM with universal high-scale supersymmetry breaking by
introducing additional SM vector-like fermions.  To avoid the
dimension-6 proton decay problem and quantum gravity effects, we
require that the gauge coupling unification scale is from $5\times
10^{15}$ GeV to the Planck scale.  We assume that the supersymmetry
breaking scale is below the unification scale, and that the universal
masses of the vector-like fermions are from 200 GeV to 1 TeV.  In
order to have the canonical gauge coupling unification and to satisfy
these requirements and assumptions, we showed that $\Delta b_2 =
\Delta b_3$ and $2/5 \le \Delta b_2 - \Delta b_1 \le 13/5$ for the
extra vector-like particles.  To systematically construct the models
with canonical gauge coupling unification, we used the technique of
the one-loop beta function equivalent relations for the particle sets.
We discussed two kinds of models. The first kind of models have
standard vector-like particles while the second kind of models have
standard and non-standard ones.  In the models with simple sets of
extra vector-like fermions whose universal masses are $200$ GeV and
$1~{\rm TeV}$, we presented the supersymmetry breaking scales, gauge
coupling unification scales, and the corresponding Higgs mass ranges.
In the first kind of models, $\Delta b_2 - \Delta b_1$ can only be
equal to $6/5$ and $12/5$, and then the corresponding supersymmetry
breaking scale can only be around $10^{10}$ GeV and $10^{15}$ GeV,
respectively.  In the second kind, $\Delta b_2 - \Delta b_1$ can
be $n/5$, in which $n=2, 3, ..., 13$, so the supersymmetry breaking
scale can be from $10^{5}$ GeV to $10^{16}$ GeV.  Because the
universal masses for the vector-like fermions are within the reach of
the LHC, these models can definitely be tested at the LHC.

We briefly commented on some phenomenological consequences
of these models, which deserve further detailed study.

\begin{acknowledgments}
  TL would like to thank H. Murayama and S. Nandi for helpful
  discussions on the gauge coupling unification in the
  non-supersymmetric models, and thank S. Thomas for useful
  conversations.  JJ thanks the National Center for Theoretical
  Sciences in Taiwan for its hospitality, where part of the work was
  done.  This research was supported by the U.S.~Department of Energy
  under Grants No.~DE-FG02-95ER40896, and DE-FG02-96ER40969, by the
  Friends of the IAS, by the Cambridge-Mitchell Collaboration in
  Theoretical Cosmology, and by the University of Wisconsin Research
  Committee with funds granted by the Wisconsin Alumni Research
  Foundation.

\end{acknowledgments}

\appendix

\section{ Renormalization Group Equations}
\label{apdxA}

In this Appendix, we give the renormalization group equations in the
SM and MSSM.  The general formulae for the renormalization group
equations in the SM are given in Refs.~\cite{mac,Cvetic:1998uw}, and
those for the supersymmetric models in
Refs.~\cite{Barger:1992ac,Barger:1993gh,Martin:1993zk}.

First, we summarize the renormalization group equations in the SM.
The two-loop equations for the gauge couplings are
\begin{eqnarray}
(4\pi)^2\frac{d}{dt}~ g_i=g_i^3b_i &+&\frac{g_i^3}{(4\pi)^2}
\left[ \sum_{j=1}^3B_{ij}g_j^2-\sum_{\alpha=u,d,e}d_i^\alpha
{\rm Tr}\left( h^{\alpha \dagger}h^{\alpha}\right) \right] ~,~\,
\label{SMgauge}
\end{eqnarray}
where $t=\ln  \mu$ and $ \mu$ is the renormalization scale.
$g_1$, $g_2$ and $g_3$ are the gauge couplings
for $U(1)_Y$, $SU(2)_L$ and $SU(3)_C$, respectively,
where we use the $SU(5)$ normalization $g_1^2 \equiv (5/3)g_Y^{ 2}$.
The beta-function coefficients are  
\begin{eqnarray}
&&b=\left(\frac{41}{10},-\frac{19}{6},-7\right) ~,~
B=\pmatrix{\frac{199}{50}&
\frac{27}{10}&\frac{44}{5}\cr \frac{9}{10} & \frac{35}{6}&12 \cr
\frac{11}{10}&\frac{9}{2}&-26} ~,~\\
&&d^u=\left(\frac{17}{10},\frac{3}{2},2\right) ~,~
d^d=\left(\frac{1}{2},\frac{3}{2},2\right) ~,~
d^e=\left(\frac{3}{2},\frac{1}{2},0\right) ~.~\,
\end{eqnarray}

Since the contributions in Eq.~(\ref{SMgauge}) from the Yukawa
couplings arise from the two-loop diagrams, we only need Yukawa
coupling evolution at one-loop order.  The one-loop renormalization
group equations for the Yukawa couplings are
\begin{eqnarray}
(4\pi)^2\frac{d}{dt}~h^u&=&h^u\left( -\sum_{i=1}^3c_i^ug_i^2
+\frac{3}{2}
h^{u \dagger}h^{u}
-\frac{3}{2}
h^{d \dagger}h^{d}
+\Delta_2\right) ~,~\\
(4\pi)^2\frac{d}{dt}~h^d&=&h^d\left( -\sum_{i=1}^3c_i^dg_i^2
-\frac{3}{2}
h^{u \dagger}h^{u}
+\frac{3}{2}
h^{d \dagger}h^{d}
+\Delta_2 \right)~,~\\
(4\pi)^2\frac{d}{dt}~h^e&=&h^e\left( -\sum_{i=1}^3c_i^eg_i^2
+\frac{3}{2}
h^{e \dagger}h^{e}
+ \Delta_2 \right) ~,~\,
\label{SMY}
\end{eqnarray}
where $h^u$, $h^d$ and $h^e$ are the Yukawa couplings
for the up-type quark, down-type quark, and lepton,
respectively. Also, $c^u$, $c^d$, and $c^e$ are given by 
\begin{eqnarray}
c^u=\left( \frac{17}{20}, \frac{9}{4}, 8\right) ~,~
c^d=\left( \frac{1}{4}, \frac{9}{4}, 8\right) ~,~
c^e=\left( \frac{9}{4}, \frac{9}{4}, 0\right) ~,~
\end{eqnarray}
and
\begin{eqnarray}
\Delta_2 &=& {\rm Tr} ( 3h^{u \dagger}h^{u}+3 h^{d \dagger}h^{d}+
h^{e \dagger}h^{e})  ~.~
\end{eqnarray}

The one-loop renormalization group equation for the Higgs quartic coupling is 
\begin{eqnarray}
(4\pi)^2\frac{d}{dt}~\lambda  &=&12 \lambda^2
-\left({9\over 5} g_1^2 + 9 g_2^2 \right) \lambda
+{9\over 4} \left( {3\over {25}} g_1^4 
+ {2\over 5} g_1^2g_2^2 + g_2^4 \right)
+4\Delta_2 \lambda - 4 \Delta_4 ~,~\,
\end{eqnarray}
where
\begin{eqnarray}
\Delta_4 &=& {\rm Tr} \left[ 3 (h^{u \dagger}h^{u})^2+3 (h^{d \dagger}h^{d})^2
+ (h^{e \dagger}h^{e})^2\right]  ~.~
\end{eqnarray}

Next, we summarize the renormalization group equations in 
the MSSM.
The two-loop renormalization group equations for the gauge couplings are
\begin{eqnarray}
(4\pi)^2\frac{d}{dt}~ g_i &=& g_i^3 b_i 
 +\frac{g_i^3}{(4\pi)^2}
\left[~ \sum_{j=1}^3 B_{ij}  g_j^2-\sum_{\alpha=u,d,e}d_i^\alpha
{\rm Tr}\left( y^{\alpha \dagger}y^{\alpha}\right) \right] ~,~\,
\label{SUSYgauge}
\end{eqnarray}
where the beta-function coefficients are 
\begin{eqnarray}
&&b=\left(\frac{33}{5},1,-3\right) ~,~~~  B=\pmatrix{\frac{199}{25}&
\frac{27}{5}&\frac{88}{5}\cr \frac{9}{5} & 25&24 \cr
\frac{11}{5}&9&14} ~,~\\
&&d^u=\left(\frac{26}{5},6,4\right) ~,~
d^d=\left(\frac{14}{5},6,4\right) ~,~
d^e=\left(\frac{18}{5},2,0\right) ~.~ \\
\end{eqnarray}

The one-loop renormalization group equations for Yukawa couplings are
\begin{eqnarray}
(4\pi)^2\frac{d}{dt}~y^u&=& y^u
\left[ 3 y^{u \dagger} y^{u}+ y^{d \dagger} y^{d}
+3{\rm Tr}( y^{u \dagger} y^{u}) 
-\sum_{i=1}^3c_i^ug_i^2 \right]~,~\\
(4\pi)^2\frac{d}{dt}~y^d&=& y^d
\left[ y^{u \dagger} y^{u} + 3 y^{d \dagger} y^{d}
+{\rm Tr}(3 y^{d \dagger} y^{d}
+ y^{e \dagger} y^{e}) 
-\sum_{i=1}^3c_i^dg_i^2 \right]~,~\\
(4\pi)^2\frac{d}{dt}~y^e&=& y^e
\left[ 3 y^{e \dagger} y^{e}+{\rm Tr}(3 y^{d \dagger} y^{d}
+ y^{e \dagger} y^{e}) 
-\sum_{i=1}^3c_i^eg_i^2 \right] ~,~\,
\end{eqnarray}
where  $y^u$, $y^d$ and $y^e$ are the Yukawa couplings
for the up-type quark, down-type quark, and lepton,
respectively. $c^u$, $c^d$, and $c^e$ are given by 
\begin{eqnarray}
&& c^u=\left( \frac{13}{15}, 3, \frac{16}{3}\right) ~,~
c^d=\left( \frac{7}{15}, 3, \frac{16}{3}\right) ~,~
c^e=\left( \frac{9}{5}, 3, 0\right) ~.~\, 
\end{eqnarray}

\section{Two-Loop Beta Functions for the Vector-Like Particles}
\label{apdxB}

In this Appendix, we present two-loop beta functions contributions
to the SM gauge couplings from the vector-like particles
which are introduced in Sections III and IV.
The general formulae are also given 
in Refs.~\cite{mac,Cvetic:1998uw,Barger:1992ac,Barger:1993gh,Martin:1993zk}.

The two-loop beta functions ($\Delta B_{ij}$) from the extra particles
in the non-supersymmetric models are
\begin{eqnarray}
\Delta B^{XQ + {\overline{XQ}}}=\pmatrix{\frac{1}{150}&
\frac{3}{10}&\frac{8}{15}\cr \frac{1}{10} & \frac{49}{2}& 8 \cr
\frac{1}{15}& 3 & \frac{76}{3} } ~,~
\Delta B^{XU + {\overline{XU}}}=\pmatrix{\frac{64}{75}&
0 &\frac{64}{15}\cr 0 & 0 & 0 \cr
\frac{8}{15} & 0 & \frac{38}{3}} ~,~\,
\end{eqnarray}
\begin{eqnarray}
\Delta B^{XD + {\overline{XD}}}=\pmatrix{\frac{4}{75}&
0 &\frac{16}{15}\cr 0 & 0 & 0 \cr
\frac{2}{15} & 0 & \frac{38}{3}} ~,~
\Delta B^{XL + {\overline{XL}}}=\pmatrix{\frac{9}{50}&
\frac{9}{10}& 0 \cr \frac{3}{10} & \frac{49}{6}& 0 \cr
0 & 0 & 0 } ~,~\,
\end{eqnarray}
\begin{eqnarray}
\Delta B^{XE + {\overline{XE}}}=\pmatrix{\frac{36}{25}&
0 & 0 \cr 0 & 0 & 0 \cr
0 & 0 & 0 } ~,~
\Delta B^{XG}=\pmatrix{ 0 &
0 & 0 \cr 0 & 0 & 0 \cr
0 & 0 & 48 } ~,~\,
\end{eqnarray}
\begin{eqnarray}
\Delta B^{XW}=\pmatrix{ 0 &
0 & 0 \cr 0 & \frac{64}{3} & 0 \cr
0 & 0 & 0 } ~,~
\Delta B^{XT + {\overline{XT}}}=\pmatrix{\frac{108}{25}&
\frac{72}{5}& 0 \cr \frac{24}{5} & \frac{128}{3}& 0 \cr
0 & 0 & 0 } ~,~\,
\end{eqnarray}
\begin{eqnarray}
\Delta B^{XS + {\overline{XS}}}=\pmatrix{\frac{128}{75}&
0 &\frac{64}{3}\cr 0 & 0 & 0 \cr
\frac{8}{3} & 0 & \frac{250}{3}} ~,~
\Delta B^{XY + {\overline{XY}}}=\pmatrix{\frac{25}{6}&
\frac{15}{2}&\frac{40}{3}\cr \frac{5}{2} & \frac{49}{2}& 8 \cr
\frac{5}{3}& 3 & \frac{76}{3} } ~,~\,
\end{eqnarray}
\begin{eqnarray}
\Delta B^{XQ0 + {\overline{XQ0}}}=\pmatrix{0&
0 & 0 \cr 0 & \frac{49}{2}& 8 \cr
0 & 3 & \frac{76}{3} } ~,~
\Delta B^{XD0 + {\overline{XD0}}}=\pmatrix{0 &
0 & 0 \cr 0 & 0 & 0 \cr
0 & 0 & \frac{38}{3}} ~,~\,
\end{eqnarray}
\begin{eqnarray}
\Delta B^{XL0 + {\overline{XL0}}}=\pmatrix{0 &
0 & 0 \cr 0 & \frac{49}{6}& 0 \cr
0 & 0 & 0 } ~,~
\Delta B^{XS + {\overline{XS}}}=\pmatrix{0 &
0 & 0 \cr 0 & 0 & 0 \cr
 0 & 0 & \frac{250}{3}} ~.~\,
\end{eqnarray}

In the supersymmetric models
\begin{eqnarray}
\Delta B^{XQ + {\overline{XQ}}}=\pmatrix{\frac{1}{75}&
\frac{3}{5}&\frac{16}{15}\cr \frac{1}{5} & 21 & 16 \cr
\frac{2}{15}& 6 & \frac{68}{3} } ~,~
\Delta B^{XU + {\overline{XU}}}=\pmatrix{\frac{128}{75}&
0 &\frac{128}{15}\cr 0 & 0 & 0 \cr
\frac{16}{15} & 0 & \frac{34}{3}} ~,~\,
\end{eqnarray}
\begin{eqnarray}
\Delta B^{XD + {\overline{XD}}}=\pmatrix{\frac{8}{75}&
0 &\frac{32}{15}\cr 0 & 0 & 0 \cr
\frac{4}{15} & 0 & \frac{34}{3}} ~,~
\Delta B^{XL + {\overline{XL}}}=\pmatrix{\frac{9}{25}&
\frac{9}{5}& 0 \cr \frac{3}{5} & 7 & 0 \cr
0 & 0 & 0 } ~,~\,
\end{eqnarray}
\begin{eqnarray}
\Delta B^{XE + {\overline{XE}}}=\pmatrix{\frac{72}{25}&
0 & 0 \cr 0 & 0 & 0 \cr
0 & 0 & 0 } ~,~
\Delta B^{XG}=\pmatrix{ 0 &
0 & 0 \cr 0 & 0 & 0 \cr
0 & 0 & 54 } ~,~\,
\end{eqnarray}
\begin{eqnarray}
\Delta B^{XW}=\pmatrix{ 0 &
0 & 0 \cr 0 & 24 & 0 \cr
0 & 0 & 0 } ~,~
\Delta B^{XT + {\overline{XT}}}=\pmatrix{\frac{216}{25}&
\frac{144}{5}& 0 \cr \frac{48}{5} & 48 & 0 \cr
0 & 0 & 0 } ~,~\,
\end{eqnarray}
\begin{eqnarray}
\Delta B^{XS + {\overline{XS}}}=\pmatrix{\frac{256}{75}&
0 &\frac{128}{3}\cr 0 & 0 & 0 \cr
\frac{16}{3} & 0 & \frac{290}{3}} ~,~
\Delta B^{XY + {\overline{XY}}}=\pmatrix{\frac{25}{3}&
15 &\frac{80}{3}\cr 5 & 21 & 16 \cr
\frac{10}{3}& 6 & \frac{68}{3} } ~,~\,
\end{eqnarray}
\begin{eqnarray}
\Delta B^{XQ0 + {\overline{XQ0}}}=\pmatrix{0 &
 0 & 0 \cr 0 & 21 & 16 \cr
 0 & 6 & \frac{68}{3} } ~,~
\Delta B^{XD0 + {\overline{XD0}}}=\pmatrix{0 &
0 & 0 \cr 0 & 0 & 0 \cr
0 & 0 & \frac{34}{3}} ~,~\,
\end{eqnarray}
\begin{eqnarray}
\Delta B^{XL0 + {\overline{XL0}}}=\pmatrix{0 &
0 & 0 \cr 0 & 7 & 0 \cr
0 & 0 & 0 } ~,~
\Delta B^{XS0 + {\overline{XS0}}}=\pmatrix{0 &
0 & 0 \cr 0 & 0 & 0 \cr
0 & 0 & \frac{290}{3}} ~.~\,
\end{eqnarray}


\begin{thebibliography}{99}
\itemsep 0.5mm


\bibitem{String}
R.~Bousso and J.~Polchinski,
JHEP {\bf 0006}, 006 (2000);
  S.~B.~Giddings, S.~Kachru and J.~Polchinski,
  Phys.\ Rev.\ D {\bf 66}, 106006 (2002);
S.~Kachru, R.~Kallosh, A.~Linde and S.~P.~Trivedi,
Phys.\ Rev.\ D {\bf 68}, 046005 (2003);
L.~Susskind,
arXiv:hep-th/0302219;
  F.~Denef and M.~R.~Douglas,
  JHEP {\bf 0405}, 072 (2004);
  JHEP {\bf 0503}, 061 (2005).


\bibitem{Weinberg}
S.~Weinberg,
Phys.\ Rev.\ Lett.\  {\bf 59} 2607 (1987).


\bibitem{Agrawal:1998xa}
  V.~Agrawal, S.~M.~Barr, J.~F.~Donoghue and D.~Seckel,
  Phys.\ Rev.\ Lett.\  {\bf 80}, 1822 (1998);
Phys.\ Rev.\ D {\bf 57}, 5480 (1998).


\bibitem{Donoghue:2003vs}
J.~F.~Donoghue,
Phys.\ Rev.\ D {\bf 69}, 106012 (2004)
[Erratum-ibid.\ D {\bf 69}, 129901 (2004)].


\bibitem{PQ} 
R. D. Peccei and H. R. Quinn, Phys. Rev. Lett. {\bf 38} 1440 (1977); 
 Phys. Rev. {\bf D16} 1791 (1977).



\bibitem{Babu:2002ic}
K.~S.~Babu, I.~Gogoladze and K.~Wang,
Phys.\ Lett.\ B {\bf 560}, 214 (2003).


\bibitem{Barger:2004sf}
  V.~Barger, C.~W.~Chiang, J.~Jiang and T.~Li,
  Nucl.\ Phys.\ B {\bf 705}, 71 (2005).

\bibitem{MGJS}
M. B. Green and J. H. Schwarz, Phys. Lett. {\bf B149} 117 (1984);
Nucl. Phys. {\bf B255} 93 (1985); M. B. Green, J. H. Schwarz and P.
West, Nucl. Phys. {\bf B254} 327 (1985).

\bibitem{Svrcek:2006yi}
  P.~Svrcek and E.~Witten,
  JHEP {\bf 0606} (2006) 051
  [arXiv:hep-th/0605206].

\bibitem{HSUSY}
A.~Giryavets, S.~Kachru and P.~K.~Tripathy,
JHEP {\bf 0408} 002 (2004);
L.~Susskind,
arXiv:hep-th/0405189;
M.~R.~Douglas,
arXiv:hep-th/0405279; arXiv:hep-th/0409207;
M.~Dine, E.~Gorbatov and S.~Thomas,
arXiv:hep-th/0407043;
E.~Silverstein,
arXiv:hep-th/0407202;
J.~P.~Conlon and F.~Quevedo,
  JHEP {\bf 0410}, 039 (2004);
  M.~Dine, D.~O'Neil and Z.~Sun,
  JHEP {\bf 0507}, 014 (2005).


\bibitem{NASD}
 N.~Arkani-Hamed and S.~Dimopoulos,
  JHEP {\bf 0506}, 073 (2005).


\bibitem{Camara:2004jj}
  P.~G.~Camara, L.~E.~Ibanez and A.~M.~Uranga,
  Nucl.\ Phys.\ B {\bf 708}, 268 (2005);
L.~E.~Ib\'a\~nez, Phys.\ Rev.\ D {\bf 71}, 055005 (2005);
  A.~Font and L.~E.~Ibanez,
  JHEP {\bf 0503}, 040 (2005).

\bibitem{LRS}
D.~L\"ust, S.~Reffert and S.~Stieberger,
  Nucl.\ Phys.\ B {\bf 727}, 264 (2005); Nucl.\ Phys.\ B {\bf 706}, 3 (2005).


\bibitem{Langacker:1991an}
  H.~Georgi and S.~L.~Glashow,
  Phys.\ Rev.\ Lett.\  {\bf 32}, 438 (1974);
  P.~Langacker and M.~X.~Luo,
  Phys.\ Rev.\ D {\bf 44}, 817 (1991);
  J.~R.~Ellis, S.~Kelley and D.~V.~Nanopoulos,
  Phys.\ Lett.\ B {\bf 260}, 131 (1991);
  U.~Amaldi, W.~de Boer and H.~Furstenau,
  Phys.\ Lett.\ B {\bf 260}, 447 (1991).
    
\bibitem{Frampton:1983sh}
  P.~H.~Frampton and S.~L.~Glashow,
  Phys.\ Lett.\ B {\bf 131}, 340 (1983)
  [Erratum-ibid.\ B {\bf 135}, 515 (1984)].
  S.~Nandi,
  Phys.\ Lett.\ B {\bf 142}, 375 (1984);
  H.~Murayama and T.~Yanagida,
  Mod.\ Phys.\ Lett.\ A {\bf 7}, 147 (1992);
  T.~G.~Rizzo,
  Phys.\ Rev.\ D {\bf 45}, 3903 (1992);
G.~F.~Giudice and A.~Romanino,
  Nucl.\ Phys.\ B {\bf 699}, 65 (2004)
  [Erratum-ibid.\ B {\bf 706}, 65 (2005)].



\bibitem{Choudhury:2001hs}
  D.~Choudhury, T.~M.~P.~Tait and C.~E.~M.~Wagner,
  Phys.\ Rev.\ D {\bf 65}, 053002 (2002).


\bibitem{Morrissey:2003sc}
  D.~E.~Morrissey and C.~E.~M.~Wagner,
  Phys.\ Rev.\ D {\bf 69}, 053001 (2004).

\bibitem{Kehagias:2005vz}
  A.~Kehagias and N.~D.~Tracas,
  arXiv:hep-ph/0506144.

\bibitem{Barger:2005gn}
  V.~Barger, J.~Jiang, P.~Langacker and T.~Li,
  Phys.\ Lett.\ B {\bf 624}, 233 (2005);
Nucl.\ Phys.\ B {\bf 726}, 149 (2005).

\bibitem{Yao:2006px}
  W.~M.~Yao {\it et al.}  [Particle Data Group],
  J.\ Phys.\ G {\bf 33}, 1 (2006).

\bibitem{Cvetic:1997ky}
  M.~Cvetic, D.~A.~Demir, J.~R.~Espinosa, L.~L.~Everett and P.~Langacker,
  Phys.\ Rev.\ D {\bf 56}, 2861 (1997)
  [Erratum-ibid.\ D {\bf 58}, 119905 (1998)]
  [arXiv:hep-ph/9703317].

\bibitem{LMN}
T.~Li, H.~Murayama and S.~Nandi, unpublished.

\bibitem{Dienes:1996du}
  K.~R.~Dienes,
  Phys.\ Rept.\  {\bf 287}, 447 (1997).


\bibitem{Blumenhagen:2005mu}
  R.~Blumenhagen, M.~Cvetic, P.~Langacker and G.~Shiu,
  Ann.\ Rev.\ Nucl.\ Part.\ Sci.\  {\bf 55}, 71 (2005).


\bibitem{Davoudiasl:2004be}
 H.~Davoudiasl, R.~Kitano, T.~Li and H.~Murayama,
  Phys.\ Lett.\ B {\bf 609}, 117 (2005).


\bibitem{Seesaw}
  H.~Fritzsch and P.~Minkowski,
  Phys.\ Lett.\ B {\bf 62}, 72 (1976);
M. Gell-Mann, P. Ramond, and R. Slansky, in {\it
Supergravity}, ed. F. van Nieuwenhuizen and D. Freedman, (North
Holland, Amsterdam, 1979) p. 315; T. Yanagida, {\it Proc. of the
Workshop on Unified Theory and the Baryon Number of the Universe},
KEK, Japan, 1979; S. Weinberg, 
Phys.\ Rev.\ Lett.\  {\bf 43}, 1566 (1979);
R. N. Mohapatra and G. Senjanovic, Phys. Rev. Lett. {\bf
44}, 912 (1980).

\bibitem{Fukugita:1986hr}
M.~Fukugita and T.~Yanagida,
Phys.\ Lett.\ B {\bf 174}, 45 (1986);
P.~H.~Frampton, S.~L.~Glashow and T.~Yanagida,
Phys.\ Lett.\ B {\bf 548}, 119 (2002);
V.~Barger, D.~A.~Dicus, H.~J.~He and T.~Li,
Phys.\ Lett.\ B {\bf 583}, 173 (2004).

\bibitem{Barger:2006fm}
  V.~Barger, J.~Jiang, P.~Langacker and T.~Li,
  arXiv:hep-ph/0612206.


\bibitem{Gogoladze:2006ps}
  I.~Gogoladze, T.~Li and Q.~Shafi,
  arXiv:hep-ph/0602040.


\bibitem{Arason:1991ic}
H.~Arason, D.~J.~Castano, B.~Keszthelyi, S.~Mikaelian, E.~J.~Piard, P.~Ramond and B.~D.~Wright,
Phys.\ Rev.\ D {\bf 46}, 3945 (1992);
H.~E.~Haber, R.~Hempfling and A.~H.~Hoang,
Z.\ Phys.\ C {\bf 75}, 539 (1997).


\bibitem{:2005cc}
    [CDF Collaboration],
  arXiv:hep-ex/0507091.

\bibitem{Brubaker:2006xn}
  E.~Brubaker {\it et al.}  [Tevatron Electroweak Working Group],
  arXiv:hep-ex/0608032.


\bibitem{Bethke:2006ac}
  S.~Bethke,
  arXiv:hep-ex/0606035.

\bibitem{Lee:2002sa}
  I.~T.~Lee {\it et al.},
  Phys.\ Rev.\ D {\bf 66}, 012002 (2002)
  [arXiv:hep-ex/0204003];
M.~L.~Perl, E.~R.~Lee and D.~Loomba,
  Mod.\ Phys.\ Lett.\ A {\bf 19}, 2595 (2004).

\bibitem{Kudo:2001ie}
  A.~Kudo and M.~Yamaguchi,
  Phys.\ Lett.\ B {\bf 516}, 151 (2001)
  [arXiv:hep-ph/0103272].

\bibitem{Gogoladze:2006si}
  I.~Gogoladze, T.~Li, V.~N.~Senoguz and Q.~Shafi,
  Phys.\ Rev.\ D {\bf 74}, 126006 (2006)
  [arXiv:hep-ph/0608181].




\bibitem{Pietroni}
M.~Pietroni, Nucl. Phys. {\bf B402}, 27 (1993).

\bibitem{DFMHS}
 A.~T.~Davies, C.~D.~Froggatt and R.~G.~Moorhouse,
Phys.\ Lett.\ B {\bf 372}, 88 (1996);
S. J.~Huber and M. G.~Schmidt,
 Eur. Phys. J. {\bf C10}, 473  (1999);
Nucl.\ Phys.\ B {\bf 606}, 183 (2001).

\bibitem{Kang:2004pp}
  J.~Kang, P.~Langacker, T.~Li and T.~Liu,
  Phys.\ Rev.\ Lett.\  {\bf 94}, 061801 (2005).


\bibitem{Braun:2005ux}
  V.~Braun, Y.~H.~He, B.~A.~Ovrut and T.~Pantev,
  Phys.\ Lett.\ B {\bf 618}, 252 (2005).


\bibitem{Orbifold}
Y.~Kawamura,
Prog.\ Theor.\ Phys.\  {\bf 103}, 613 (2000);
G. Altarelli and F. Feruglio, 
Phys.\ Lett.\ B {\bf 511}, 257 (2001);
L. Hall and Y. Nomura, Phys.\ Rev.\ D {\bf 64}, 055003 (2001);
A. Hebecker and J. March-Russell, 
Nucl.\ Phys.\ B {\bf 613}, 3 (2001);
 T. Li, Phys.\ Lett.\ B {\bf 520}, 377 (2001);
Nucl.\ Phys.\ B {\bf 619}, 75 (2001).


\bibitem{Barger:1985nq}
  V.~D.~Barger, N.~Deshpande, R.~J.~N.~Phillips and K.~Whisnant,
  Phys.\ Rev.\ D {\bf 33}, 1912 (1986)
  [Erratum-ibid.\ D {\bf 35}, 1741 (1987)].

\bibitem{Hewett:1988xc}
  J.~L.~Hewett and T.~G.~Rizzo,
  Phys.\ Rept.\  {\bf 183}, 193 (1989).

\bibitem{KLN}
  J. Kang, P. Langacker, and B. D. Nelson, in preparation.

\bibitem{Fairbairn:2006gg}
  M.~Fairbairn, A.~C.~Kraan, D.~A.~Milstead, T.~Sjostrand, P.~Skands and T.~Sloan,
  arXiv:hep-ph/0611040.

\bibitem{Barger:1995dd}
  V.~D.~Barger, M.~S.~Berger and R.~J.~N.~Phillips,
  Phys.\ Rev.\ D {\bf 52}, 1663 (1995)
  [arXiv:hep-ph/9503204].


\bibitem{Deshpande:2003nx}
  N.~G.~Deshpande and D.~K.~Ghosh,
  Phys.\ Lett.\ B {\bf 593}, 135 (2004).


\bibitem{REST}
R. Essig and S. Thomas, in preparation.


\bibitem{Carena:2004ha}
  M.~Carena, A.~Megevand, M.~Quiros and C.~E.~M.~Wagner,
  Nucl.\ Phys.\ B {\bf 716}, 319 (2005).


\bibitem{Arvanitaki:2005fa}
  A.~Arvanitaki, C.~Davis, P.~W.~Graham, A.~Pierce and J.~G.~Wacker,
  Phys.\ Rev.\ D {\bf 72}, 075011 (2005).


\bibitem{Grossman:2003gq}
Y.~Grossman and S.~Rakshit,
Phys.\ Rev.\ D {\bf 69}, 093002 (2004).


\bibitem{mac}
M.~E.~Machacek and M.~T.~Vaughn,
Nucl.\ Phys.\ B {\bf 222}, 83 (1983);
Nucl.\ Phys.\ B {\bf 236}, 221 (1984);
Nucl.\ Phys.\ B {\bf 249}, 70 (1985).


\bibitem{Cvetic:1998uw}
G.~Cvetic, C.~S.~Kim and S.~S.~Hwang,
Phys.\ Rev.\ D {\bf 58}, 116003 (1998).

\bibitem{Barger:1992ac}
V.~D.~Barger, M.~S.~Berger and P.~Ohmann,
Phys.\ Rev.\ D {\bf 47}, 1093 (1993).

\bibitem{Barger:1993gh}
V.~D.~Barger, M.~S.~Berger and P.~Ohmann,
Phys.\ Rev.\ D {\bf 49}, 4908 (1994).


\bibitem{Martin:1993zk}
S.~P.~Martin and M.~T.~Vaughn,
Phys.\ Rev.\ D {\bf 50}, 2282 (1994), and references therein.


\end{thebibliography}
\end{document}